\documentstyle[epsf,pra,aps]{revtex}

\begin{document}
\draft

\def\del{\partial}
\def\eq{\begin{equation}}
\def\eeq{\end{equation}}
\def\eqa{\begin{eqnarray}}
\def\eeqa{\end{eqnarray}}
\def\nn{\nonumber}

\title{Plateau transitions in fractional quantum Hall liquids}
\author{Ken-Ichiro Imura
\footnote{Former address: Department of Applied Physics,
University of Tokyo, 113-8656, Tokyo.}}
\address{Laboratoire de Physique Theorique et Modeles Statistiques\\
Batiment 100, Universite Paris-Sud, 91405 Orsay Cedex, France}
\date{\today}
\maketitle

\begin{abstract}
Effects of backward scattering between fractional quantum Hall 
(FQH) edge modes are studied. Based on the edge-state picture 
for hierarchical FQH liquids, we discuss the possibility of 
the transitions between different plateaux of the tunneling 
conductance $G$. We find a selection rule for the sequence 
which begins with a conductance $G=m/(mp\pm 1)$ ($m$: integer, 
$p$: even integer) in units of $e^2/h$. The shot-noise spectrum 
as well as the scaling behavior of the tunneling current is 
calculated explicitly.
\end{abstract}

\pacs{72.10.-d, 73.20.Dx, 73.40.Hm}


\section{introduction}
The fractional quantum Hall (FQH) effect is a phenomenon observed in
a two-dimensional electron system subjected to a strong perpendicular
magnetic field.
Due to the interplay between the strong magnetic field and interactions
among the electrons as well as weak disorder, the transverse resistivity
shows a plateau behavior. \cite{tsui}
For a filling factor $\nu=1/$(odd integer), the theory predicts fractionally
charged quasiparticles with charge $q=\nu e$. \cite{bob}
Recent shot-noise experiments in a two-terminal FQH system with a
point-like constriction or a point quantum contact (QPC) between the edges
seem to be consistent with this theoretical
prediction.\cite{cglat}
The FQH system which have any experimental relevance should be confined
in a finite region enclosed by one or more edges.
Due to the presence of strong magnetic field, the
low-energy physics of this two-dimensional electron liquid reduces
essentially to that of the one-dimensional edge mode.
In one dimension it is known that the interaction plays
a significant role. The electrons are strongly renormalized
so that the Fermi liquid theory breaks down to be replaced
by the Tomonaga-Luttinger liquid (TLL).~\cite{hald} 
If one considers spinless
electrons in 1D, the TLL is completely characterized by one
parameter $g$ which represents the strength of interaction.
Therefore the parameter $g$ for an interacting electron system
in 1D is not universal. On the other hand a remarkable feature
of FQH edge mode is that the parameter $g$ which controls
this 1D system is universal, since $g$ is related to the
topological nature of the bulk FQH liquid (FQHL).
For the edge mode of principal Laughlin states, the parameter
$g$ is simply given by the bulk filling factor $\nu$.
\cite{wen1}

The edge-tunneling experiment in FQH liquids has shown
that a chiral TLL is realized at the edge of FQHL. \cite{chang}
Indeed the chiral TLL theory has succeeded in the description of
non-linear $I-V$ characteristics for $\nu=1/$(odd integer),\cite{wen2}
but it is also true that the edge-tunneling experiment 
cannot be explained by a naive TLL theory for other filling factors. 
\cite{grayson,dhlee}
In particular, for the Jain's composite fermion hierarchy states
at filling factor
$\nu=m/(mp+\chi)$ ($m$: integer, $p$: even integer, $\chi=\pm 1$),
\cite{jain}
Wen's chiral TLL theory predicts that there should be $m$
edge modes corresponding to each composite fermion Landau level.
\cite{wen3}
Due to the existence of these internal degrees of freedom
the predicted exponent $\alpha$ for the $I-V$ characteristics 
does not fit the experiment: $\alpha\sim 1/\nu$.

Although the observed exponent $\alpha\sim 1/\nu$ for the tunneling
into FQHL does not support the hierarchial structure of edge mode,
there is another experimental observation which encourages us to
work on this theory. It is the suppressed shot-noise
measurement at bulk filling factor $\nu=2/5$, i.e., at $m=2,p=2,\chi=1$
in a constricted two-terminal Hall bar geometry. \cite{rez}
They observed the transitions of two-terminal conductance 
from a plateau at $G=2/5$ to another at $G=1/3$ and finally to $G=0$
as the constriction is increased.
On the plateau at $G=1/3$ they observed a fractional charge $q=e/3$,
which indicates that the filling factor near the quantum point
contact (QPC) is $\nu=1/3$. 
The experiment clearly indicates a deep connection
between the $\nu=2/5$ daughter state and the $\nu=1/3$ parents state,
and therefore seems to support the hierachy theory at $\nu=2/5$.

This paper studies the tunneling through a QPC at the edge of
FQHL. This topic has captured a widespread attention both 
experimentally and theoretically. For the reasons stated above
we forcus on the filling factor $\nu=m/(mp+\chi)$.
The FQH systems at those filling factors will provide
an interesting arena to study the hierarchical 
nature of those liquids. 
We discuss the successive transition between the plateaux
of conductance. The sequence begins with the conductance
$G=G_m=m/(mp+\chi)$ in units of $e^2/h$, which
is identical to the bulk filling factor. We discuss
the selection rule for the transitions between different $G$'s.
Even though the eventual correctness
of the hierachical TLL theory description is yet to be
tested, on which we will be based, we insist that 
it is of importantance to make various interesting 
applications of the theory. It will enable us to compare the
experiment with the theoretical predictions, and hence
will be useful to judge the correctness of hierarchical picture.

\section{model}
Our model has a two-terminal Hall bar geometry.
The bulk FQHL in the $xy$-plane is confined electro-statically
on the $y$-direction into a finite region: $-w/2<y<w/2$.
Each end of this strip is connected to a source (the left terminal)
or to a drain (the right terminal).
We assume that the bulk FQHL is incompressible at a filling
factor $\nu$. Therefore
the low-energy excitations are allowed only in the vicinity
of two boundaries, which constitute the edge modes.
The upper (lower) edge mode carries a current from the
left (right) to the right (left) terminal, and the total
current $I$ is defined as the difference of two.
Since there is no mechanism of relaxation in the TLL itself,
the chemical potential is uniform in the respective edge modes, i.e.,
the upper (lower) edge mode has a chemical potential
equal to that of the source (drain).
All the scatterings occur inside the terminals.
\cite{land}
In the absence of point-like constriction, the two-terminal
conductance $G=I/V$ is quantized at $G=\nu$ in units of $e^2/h$,
since the back-scattering between the two edge modes
which breaks the momentum conservation is allowed
nowhere through the edge,
where $V$ is defined as the source-drain voltage.

Now we go back to the bulk FQHL. We forcus on a filling factor 
$\nu=m/(mp+\chi)$ in the Jain's composite fermion hierarchy series,
where $m$: integer, $p$: even integer and $\chi=\pm 1$.
According to the bulk hierarchy structure, there should be $m$ 
edge modes, i.e., each edge mode corresponds to a composite 
fermion Landau level in the bulk. Then
the low-energy physics of this electron liquid is
controlled by the $m$-channel edge mode, which obey the following
Lagrangian density, \cite{wen3}
\eq
{\cal L}_{\rm TLL}=
{i\over 4\pi}K^{\alpha\beta}
{\del\phi_\alpha^+\over\del\tau}
{\del\phi_\beta^-\over\del x}
+{1\over8\pi}U^{\alpha\beta}
\left(
\frac{\partial\phi_\alpha^+}{\partial x}
\frac{\partial\phi_\beta^+}{\partial x}
+
\frac{\partial\phi_\alpha^-}{\partial x}
\frac{\partial\phi_\beta^-}{\partial x}
\right),
\eeq
where $\phi^\pm=\phi^u\pm\phi^l$ with $\phi^u (\phi^l)$ being the edge mode
propagating near the upper (lower) boundary of the system.
The matrix $K$ in Eq.~(1) could be identified as
the so-called $K$-matrix in the bulk, which together with
the electromagnetic-charge vector $t$ completely classify the universal
properties of bulk FQHL.
\cite{zee}
The standard construction for the $K$-matrix at a hierarchical
filling factor $\nu=m/(mp+\chi)$
yields
\eq
K=K(m,p,\chi)=\chi I_m +pC_m
\eeq
in the unitary basis $t^T=(1, \cdots, 1)$,
where $I_m,C_m$ are $m\times m$ identity and pseudo-identity matrices.
By a linear transformation one can decompose the modes into
charge and pseudo-spin bosons.
Each row and column of the matrices corresponds to a Landau level for
the composite fermions, i.e., $\alpha, \beta=1, \cdots, m$.
However it would be fair to comment that it is still a controversial
question what the correct construction of the $K$-matrix is.
\cite{lopez}
The matrix $U$ in Eq.~(1) is a positive definite matrix,
which specifies among others the velocities of the edge modes.
For $\chi=1$ charge and pseudo-spin modes propagate in the same
direction (co-propagate),
whereas for $\chi=-1$ they are counter-propagating, i.e.,
$\chi$ stands for the chirality of the edge modes.
For the latter case ($\chi=-1$), the interaction between
the edge modes can make the conductance non-universal.
The observed conductance, on the contrary, seems to be
universal. A remedy for this puzzle would be to put disorder
along the edge.~\cite{KFP} In the presence of such disorder
our conclusions will be modified, however, which will
not be discussed in the body of the paper.

Now we introduce the back-scattering by pinching the Hall bar,
i.e., by breaking the global translational invariance at $x=0$.
Let us think of applying a gate voltage locally in the middle 
of Hall bar. It squeezes the Hall bar and makes a quantum
point contact (QPC) between the two edges.
The QPC introduces the tunneling of quasiparticle through the
pinched region of Hall bar.
In the TLL model it corresponds to a backward scattering
and hence could be described by a periodic
potential barrier for the bosonic fields.
\cite{KFFN}
Let us remember that we are focusing on the bulk filling factor
$\nu_{\rm bulk}=m/(mp+\chi)$.
According to the Jain's composite fermion hierarchy,
there should be $m$ filled composite fermion Landau levels
in the bulk, and accordingly $m$ types of elementary
quasiparticles. Each correspond to a vortex-charge vector
$l=l_j$ where $(l_j)^\alpha=\delta_j^\alpha$ ($j,\alpha =1,\cdots,m$)
with $\delta_j^\alpha$ being unity for $\alpha =j$ and vanishes
otherwise.~\cite{blok} The fractional charge carried by the 
quasiparticle $l$ is given in general as
\eq
q/e=t^T K^{-1} l = {1\over mp+\chi}\sum_{\alpha=1}^m l^\alpha.
\label{charge}
\eeq
For the elementary quasiparticles $l=l_j$ one finds
$q=e/(mp+\chi)$, which is indeed the smallest possible value.
For $m=2,p=2,\chi=1$, i.e., $q=e/5$ they could be identified
as the current-carrying particles observed in the recent
shot-noise experiment at $\nu=2/5$. \cite{rez}

The tunneling of quasiparticle on the $j$-th composite fermion
Landau level induces a potential barrier proportional to
$\delta (x)\cos\phi_j^{+}$. Since the scattering amplitude
would be different for different types of quasiparticles,
the tunneling of these elementary quasiparticles sums up to
the following scattering potential barrier:
\eq
{\cal L}_{\rm tun}^{\rm (I)}=\sum_{j=1}^m
u_j\delta (x)\cos\phi_j^{+}.
\label{I}
\eeq
We call it the scattering potential due to the tunneling
of Class [I] quasiparticles.
In Eq. (\ref{I}) we took into account only the `intra-Landau-level'
processes. 

Now I draw your attention to another class of quasiparticles,
which we call Class [II]. It consists of $m$ elementary quasiparticles.
The vortex-charge vector assigned to this Class [II] quasiparticle
is $l^T=(1,\cdots,1)$. In this combination of the bosonic fields
all the neutral modes cancel and the tunneling of such
quasiparticles do not accompany any neutral modes.
The fractional charge carried by this quasiparticle is
found to be $q/e=m/(mp+\chi)=\nu$.
In the bosonic language it can be written as
$\phi_c=\phi_1+\cdots+\phi_m$, which indeed corresponds to
the charge mode.
The scattering potential due to the Class [II] quasiparticle
tunneling operator can be written as
\eq
{\cal L}_{\rm tun}^{\rm (II)}=u\delta (x)\cos\phi_c^{+}.
\label{II}
\eeq
Now our total Lagrangian density reads
${\cal L}_{\rm total}={\cal L}_{\rm TLL}+{\cal L}_{\rm tun}^{\rm (I)}
+{\cal L}_{\rm tun}^{\rm (II)}$.

In the RG analysis in Sec. IV, we study the scaling behavior
of $u_j$'s and $u$, which are controlled by the scaling dimesions
of the quasiparticle tunneling operators: $\cos\phi_j^{+}$
and $\cos\phi_c^{+}$. They are given by
$\Delta_I=\nu/m^2+1-1/m$ for the Class [I] quasiparticles,
whereas $\Delta_{II}=\nu$ for the Class [II]
quasiparticle, where $\nu=m/(mp+\chi)$.~\cite{wen2}
If the scaling dimension is smaller than 1, the corresponding
tunneling amplitude tends to have stronger values as the
voltage or the temperature is lowered.
It is indeed the case both for $u_j$'s and $u$.
In the parameter region $\{(m,p)|m\geq 2, p\geq 2\}$ in which we are
interested, one can prove that
\eq
\left\{
\begin{array}{l}
\Delta_I=\Delta_{II}\ \ \ {\rm for}\ \ \ \chi=-1, m=2, p=2\ \ \ (\nu=2/3)
\\
\Delta_I>\Delta_{II}\ \ \ {\rm otherwise}
\end{array}
\right.,
\eeq
i.e., the Class [II] quasiparticles have a lower scaling dimension
in most of the cases, and hence more relevant in the RG sense.
Another important observation is that there would be at least
two ways how the scattering becomes stronger. One way is, as we 
have described above, to increase its amplitude. However it would
be also possible that higher-order cascade of scatterings becomes 
important, where the single QPC description is no longer valid.
One might have to take into account the resonance in such 
non-perturbative regime.

\section{Hypotheses}
In Sec. II we gave expressions to the possible tunneling
processes through a single QPC in terms of the bosonic
field $\phi_j$ or of its linear combination $\phi_c$.
We considered two classes of quasiparticles. The Class [I]
corresponds to the elementary quasiparticles with the
smallest fractional charge. 
The Class [II] corresponds to the charge mode: 
$\phi_c=\phi_1+\cdots+\phi_m$.
We also compared the scaling dimensions of the two classes
of quasiparticle tunneling operators. Although the Class [II] 
quasiparticles have a lower scaling dimension in most of 
the cases and more relevant in the RG sense, 
it is not unlikely that the Class [II] 
is negligible for some reason; since they are
bound states of $m$ elementary quasiparticles,
they are so scarcely created that
the scattering potential (\ref{II}) could not develop enough to
be effective at the energy scales in question despite its
relevant scaling dimension.
Therefore we are encouraged to consider the following cases:
\begin{enumerate}
\item
Case [A]:
Class [II] quasiparticles are negligible for some reason. 
Furthermore the tunneling amplitudes for
each channel $j$ have different orders of magnitude:
\eq
u_m\gg u_{m-1}\gg\cdots\gg u_1.
\label{A}
\eeq

\item
Case [B]:
Class [II] is still negligible, but some $u_j$'s have comparable
orders of magnitude;
\eq
u_{j+1}\gg u_j\sim\cdots\sim u_{j-k+1}\gg u_{j-k},
\label{B}
\eeq
where $k\geq 2$ is an integer.

\item
Case [C]:
Class [II] is no longer negligible, i.e.,
the amplitudes $u$ for the Class [II] quasiparticle has a comparable
magnitude with those for $u_j$'s.

\end{enumerate}
The assumption (\ref{A}) for Case [A] might be justified in a way analogous
to the edge-channel argument for integer quantum Hall effect
(IQH). \cite{halp}
Let us consider the Landau levels for composite fermions.
The lowest $m$ Landau levels are completely filled by the
composite fermions, and the chemical potential lies
between the $m$-th and $(m+1)$-th Landau levels.
Towards the edge of the sample each energy level tends to be
lifted up by the confining potential.
Since the $j$-th edge mode lies in where the chemical potential crosses
the $j$-th energy level, each edge channels are spatially
separated. Therefore the tunneling between the
$m$-th edge modes, the spatially closest ones, is supposed to have
a much larger amplitude than the other $m-1$ channels.
It is also the case for $u_{m-1}$ compared with the remaining
$m-2$ channels and so forth.
I would like to mention an experiment which encourages us to
employ the assumptions (6).
It is the suppressed shot-noise
measurement at bulk filling factor $\nu=2/5$, i.e., at $m=2,p=2,\chi=1$.
~\cite{rez}
They observed the transitions of two-terminal conductance
from a plateau at $G=2/5$ to another at $G=1/3$ and finally to $G=0$
as the constriction is increased.
On the plateau at $G=1/3$ they observed a fractional charge $q=e/3$,
which indicates that the filling factor near the QPC is $\nu=1/3$.
Hence a single-channel edge mode described by a
$1\times 1$ $K$-matrix; $K=3$ is expected near the QPC.
~\cite{nom}
On the other hand in the region where the physics is completely
unaffected by the gate, the matrix $K$ should be given by
$K=K(2,2,1)$.
This experiment not only indicates a deep connection
between the $\nu=2/5$ daughter state and the $\nu=1/3$ parents state.
It also implies that $u_2\gg u_1$ as well as $u$ is negligiblly small.

We have explained above a physical reason why we are intereted
in the parameter region (\ref{A}). However we have another example
where the assumption (\ref{B}) seems to be reasonable.
It is the spin-singlet state at $\nu=2/3$
($m=2,p=2,\chi=-1$), where $j=1,2$ corresponds to each spin indices
instead of each Landau level for the spin-polarized state.
If we neglect the Zeeman energy, the theory should be symmetric
in terms of the two spin components.
Therefore it seems more reasonable to assume in this case as
$u_1\sim u_2$, which belongs to our Case [B].~\cite{rapid}

\section{RG flow and critical phenomena}
Let us forget for the moment the Class (II) quasiparticles.
Then we consider the renormalization group (RG) phase diagram in the
$m$-dimensional space of $u =(u_1, \cdots, u_m)$.
The origin of this plane
corresponds to the conductance plateau at 
$G=m/(mp+\chi)=\nu_{\rm bulk}$.
A standard RG analysis shows that
only the fixed point at $u=(\infty,\cdots,\infty)$,
is infra-red (IR) stable,
since all $u_j$'s are found to be relevant.
We introduce a small negative gate voltage to the system
on the conductance plateau at $G=G_m$.
We fix the gate voltage
so that the scaling at the zero temperature
should be controlled by the voltage difference $qV$
between the two reservoirs. Now we ask where the initial 
point of our RG flow is.

\subsection{Successive transistions}
Let us consider the Case [A]. Our RG flow starts from a point 
in the vicinity of the origin (unstable fixed point)
where Eq. (\ref{A}): $u_m\gg u_{m-1}\gg\cdots\gg u_1$ is satisfied.
As the voltage is decreased, all $u_j$'s scale to
larger values. But due to the assumption (\ref{A})
$u_m$ increases much faster than the other $m-1$ channels.
Therefore our RG path flows into the domain $D_{m-1}$, where
$D_j$ ($j=1,\cdots,m-1$) is defined as
\eq
D_{j}=\{(u_1,\cdots,u_m)|
u_m,\cdots,u_{j+1}>u_{\rm crtc}\gg u_{j},\cdots,u_1\}.
\eeq
$u_{\rm crtc}$ is a critical value of the tunneling
amplitudes such that the phase $\phi_j$ is pinned
when $u_j>u_{\rm crtc}$
in order for $g$ to be quantized.
In reality $u_{crtc}$ is determined by the strength of impurity
potential which could retain the induced quasiparticles
at the impurity cite.
In the domain $D_{m-1}$ the effective $K$-matrix near the PC
reduces to $K=K(m-1,p,\chi)$.
Therefore one could indentify $D_{m-1}$
to the plateau of conductance at $G=G_{m-1}$.
However since the domain $D_{m-1}$ corresponds
to a saddle region of the RG flow,
our RG path flows away from $D_{m-1}$
and goes toward the next saddle region $D_{m-2}$
defined in the same way as $D_{m-1}$.
We further introduce the domain $D_j$ in general for
$j=1,\cdots,m-1$.
Our RG flow passes through $D_j$'s as
\eq
D_{m-1}\rightarrow D_{m-2}\rightarrow\cdots\rightarrow D_1,
\eeq
and finally it flows into the attractive
fixed point $u=(\infty,\cdots,\infty)$,
which is identified as the completely reflecting phase $G=0$:
the domain $D_0$. (Fig. 1)
The exceptions are the series belonging to $\chi=-1, p=2$,
i.e., the $\nu=2/3$ state and its daghter states.
For those filling factors our RG stops at the
$G=1$ plateau.~\cite{nom,PRB}
In the following we consider the other cases.
As the RG path flows from the vicinity of the origin toward
the $G=0$ phase, the effective $K$-matrix near the PC
changes as
\eq
K(m,p,\chi)\rightarrow K(m-1,p,\chi)\rightarrow \cdots \rightarrow 
K(1,p,\chi)=p+\chi \rightarrow {\rm insulator}.
\eeq
Correspondingly we predict the following successive
plateau transitions,
\eq
G_m\rightarrow G_{m-1}\rightarrow\cdots\rightarrow G_1={1\over p+\chi}
\rightarrow 0.
\label{cond}
\eeq
One might think the above result is very close to
the `global phase diagram' in the quantum Hall effect.~\cite{KLZ}
Though it indeed is, it differs in that
the direction of the transition is specified in
our case.
Anomalous transitions ($G_j\rightarrow G_{j-k}$ for $k\geq 2$)
are forbidden as far as the assumption (6) is satified.

I would like to deduce the scaling behavior of the
tunneling current and the shot-noise spectrum on the plateau $G=G_j$
($j=1,\cdots,m$).
The back-scattering current
$I_b=\nu (e^2/h)V-I$
can be calculated perturbatively with respect to $u_{j-1}$
and obtained as~\cite{wen1}
\eq
\langle I_b \rangle
={2\pi q\over \Gamma[2\Delta_I]}|u_{j-1}|^2
{a^{2\Delta_I-2}\over v_c^{2\nu/m^2}v_s^{2(1-1/m)}} (qV)^{2\Delta_I -1},
\eeq
where $q=e/(jp+\chi)$ is a fractional charge of the
elementary quasiparticle on the plateau $G=G_j$, and
$\Delta_I$ is a scaling dimension of the Type (I) quasiparticle tunneling
operator: $\Delta_I=\nu/m^2+1-1/m$.
$a$ is a short-distance cutoff, and $v_c$ and $v_s$ are velocities
of the charge and the pseudo-spin modes, respectively.
The shot-noise spectrum
\eq
S (\omega)=\int_{-\infty}^{\infty}dt\cos\omega t
\langle\{I_b(t),I_b(0)\}\rangle
\eeq
is also calculated perturbatively to give,
$S (\omega)=q \langle I_b \rangle
(|1-{\omega/qV}|^{2\Delta_I-1}+
 |1+{\omega/qV}|^{2\Delta_I-1})$,
which reduces to $S=2q \langle I_b \rangle$
in the white-noise limit ($|\omega|\ll qV$).~\cite{chmn}
They are also calculated near the insulating phase to be
$S(\omega)=2e\langle I\rangle$ with
\eq
\langle I \rangle
={2\pi e\over \Gamma[2(p+\chi)]}|\tilde{u_1}|^2
a^{2(p+\chi)-2} \tilde{v_1}^{-2(p+\chi)} (eV)^{2(p+\chi)-1},
\eeq
where $\tilde{u_1}$ represents
the strength of electron tunneling dual to $u_1$
and $\tilde{v_1}$ a corresponding velocity.

\subsection{Anomalous transition} 
Let us turn to the case [B], where the Type [II] quasiparticles are
still negligible, but some $u_j$'s have comparable orders of
magnitude.
Here we consider a particular case of the assumption (7);
we consider the case where all $u_j$'s have comparable orders of
magnitude:
\eq
u_m\sim u_{m-1}\sim\cdots\sim u_1.
\label{B2}
\eeq
In this case a direct transition from $G=G_m$ to $G=0$
is expected, since all $\phi_j$'s tend to be pinned
at the same speed. The shot-noise spectrum on the plateau 
at $G=G_m$ is given by $S(\omega)=2q\langle I_b\rangle$ 
for $|\omega|\ll qV$ with
\eq
\langle I_b \rangle
={2\pi q\over \Gamma[2\Delta_I]}
\sum_{j=1}^{m}|u_{j}|^2
{a^{2\Delta_I-2}\over v_c^{2\nu/m^2}v_s^{2(1-1/m)}} (qV)^{2\Delta_I -1}.
\eeq
Now we turn our discussion to the insulating phase: $G=0$.
Remember each vortex-charge vector $l$ with integer elements
corresponds to a quasiparcile which has a charge given by
(\ref{charge}).
To construct an electron operator, we have only to set
Eq. (\ref{charge}) to be equal to 1. Of course, there is
in priciple an infinite number of choice of $l$ to
make it identical to unity.
However, as far as the tunneling is concerned,
we can pick up most relevant electron operators,
which are found to be \cite{blok}
\eq
l=\tilde{l}_j,\ \ \ (\tilde{l}_j)^\alpha=p+\chi\delta_j^\alpha,
\eeq
where $j=1,\cdots,m$ and each $\tilde{l}_j$
has $m$ components, i.e., $\alpha=1,\cdots,m$.
These electrons look analogous to our Class [I] quasiparticles.
It indeed is, but we will see that the relation is deeper.
Before going into that, the scattering potential barrier
due to the tunneling of these `Class [I] electrons'
can be written as
\eq
\tilde{{\cal L}}_{\rm tun}^{\rm (I)}=\sum_{j=1}^m
\tilde{u}_j\delta (x)
\cos\left[
\sum_{\alpha=1}^m (\tilde{l}_j)^\alpha \tilde{\phi}_\alpha^{+}
\right].
\label{ET}
\eeq
Here the `inter-Landau-level' tunnelings are neglected again.
Note that the $x$-axis is taken along the edge which is
assumed to be completely reflected in the insulating phase.

Let us take notice of the duality between
the quasiparticle tunneling and the electron tunneling,
which is exact when $u_m =u_{m-1}=\cdots =u_1$.
~\cite{rapid}
To see this, let us go back to the weak-scattering phase,
i.e., we start with the Lagrangian density:
${\cal L}_{\rm total}={\cal L}_{\rm TLL}+{\cal L}_{\rm tun}^{\rm (I)}$.
We starts our RG from the vicinity of the origin
in the $m$-dimensional space of $u =(u_1, \cdots, u_m)$.
We assume that the condition (\ref{B2}) is satisfied.
As the voltage is decreased, all $u_j$'s scale to
larger values. Then one is encouraged to employ the duality
transformation, i.e., one considers the tunneling of
instantons between the potential minima.~\cite{schmid}
Up to the lowest non-trivial order with respect to
those instantons, one obtains a model which has exactly
the same form as Eq. (\ref{ET}),
where $-\tilde{u}_j /2$ corresponds to an instanton fugacity
whereas \{$\tilde{\phi}_\alpha^{+}$\} is identified as a set of 
bosonic fields dual to \{$\phi_\alpha^{+}$\}.

The shot-noise spectrum in the insulating phase
is given by $S(\omega)=2e\langle I\rangle$ for $|\omega|\ll qV$
with the tunneling current $I$ scaling as
\eq
\langle I \rangle
={2\pi e\over \Gamma[2\tilde{\Delta}_I]}
\sum_{j=1}^{m}|\tilde{u}_j|^2
{a^{2\tilde{\Delta}_I-2}\over 
\tilde{v}_c^{2/\nu}\tilde{v}_s^{2(1-1/m)}} 
(eV)^{2\tilde{\Delta}_I-1}.
\eeq
The scaling dimension $\tilde{\Delta}_I$ of our Class [I] electron
tunneling operator is given by $\tilde{\Delta}_I=1/\nu+1-1/m$.
$\tilde{v}_c,\tilde{v}_s$ are velocities in the insulating phase.

\subsection{Direct transition}
Let us consider the case [C]. In this case we obtain
still different results.
In the presence of Class [II] quasiparticles,
the physics tends to be controlled by the scattering potential
$u$ as the energy in question is lowered.
In the region where $u_1,\cdots,u_m\ll u\ll qV$ is
satisfied, i.e., $G\sim G_m$, one obtains
$S(\omega)=2q\langle I_b\rangle$ with
\eq
\langle I_b \rangle
={2\pi q\over \Gamma[2\Delta_{II}]}|u|^2
a^{2\Delta_{II}-2} v_c^{-2\Delta_{II}} (qV)^{2\Delta_{II}-1}.
\eeq
The fractional charge of the Class [II] quasiparticle
is identical to the bulk filling factor:
$q/e=m/(mp+\chi)=\nu$, which is also equal to the scaling
dimension of the corresponding quasiparticle tunneling operator:
$\Delta_{II}=m/(mp+\chi)=\nu$.
As the RG path flows into the strong-scattering phase,
the conductance $G$ shows a direct transition to $G=0$ again.
However the scaling behavior of the tunneling current
in the insulating phase is less clear. The reason is that 
the duality is not existing in this case so that the
physical interpretation of the strong-scattering phase Lagrangian
is lacking.

\section{summary: the selection rule}
In summary we obtained the following selection rules
for the transition between plateaux starting with
the bulk value $G=m/(mp+\chi)=\nu$.
For Case [C] a direct transition to the Hall insulator
is expected.
For Cases [A] (and [B]) successive transitions from one $G$
to another $G$ is allowed under the following
selection rule:
\eqa
g=\nu(j,p,\chi)&\rightarrow&g=\nu(j',p',\chi')
\nn \\
j'=j-1,\ \ p'&=&p,\ \ \chi'=\chi.
\eeqa
An anomalous transition $j\rightarrow j-k$ ($k>2$: integer) is expected
between the same $p$ and $\chi$ when the condition (\ref{B})
is satisfied.

The overall picture of the system which
results from the plateaux transitions discussed above is the following.
We started with the situation where the filling factor
is extended uniformly over the whole system, i.e., equal to the bulk
value $\nu=m/(mp+\chi)$.
Then we effectively increased the gate voltage
in units of the voltage difference $V$ between the two terminals.
We found successive transitions of the conductance
(\ref{cond}) when the condition (\ref{A}) is satisfied.
The question is what happens
between the QPC and the bulk FQHL.
Each time $G$ passes through one plateau
($G=G_j$), there should appear one additional incompressible strip
with a filling factor $\nu=j/(jp+\chi)$.

Before ending this paper I mention that the edge-confining potential
is assumed to be steep through the paper enough to avoid the complexities
which may arise when the confining potential is smooth.
\cite{been,chkl}
In conclusion we studied the successive transitions of conductance
between different plateaux of hierarchical FQHL.
The scaling behavior of the tunneling current
and the shot-noise spectrum are calculated perturbatively
on each plateau of the conductance.
We discussed the selection rules 
for the transition between different plateaux of the conductance
in order that the theory could be tested by the experiments.

\acknowledgements
This paper is an extension and a generalization of an earlier
paper with K. Nomura (Ref.~\cite{nom}).
I am grateful to him for useful discussions and a collaboration.
I am also grateful to Y. Morita for a key comment to
initiate the present work.
I would like to thank P. Lederer and N. Nagaosa for their
suggestions and encouragements.
I would like to acknowledge the kind hospitality 
during my participance in the Trimestre Fermions Fortement 
Correles (IHP, Paris).
I was supported by JSPS Research Fellowships for Young Scientists,
and am supported by Ministere de l'Education Nationale,
de la Recherche et de la Technologie.

\begin{figure}[h]
\input epsf
\epsfxsize=15.0cm
\epsfbox{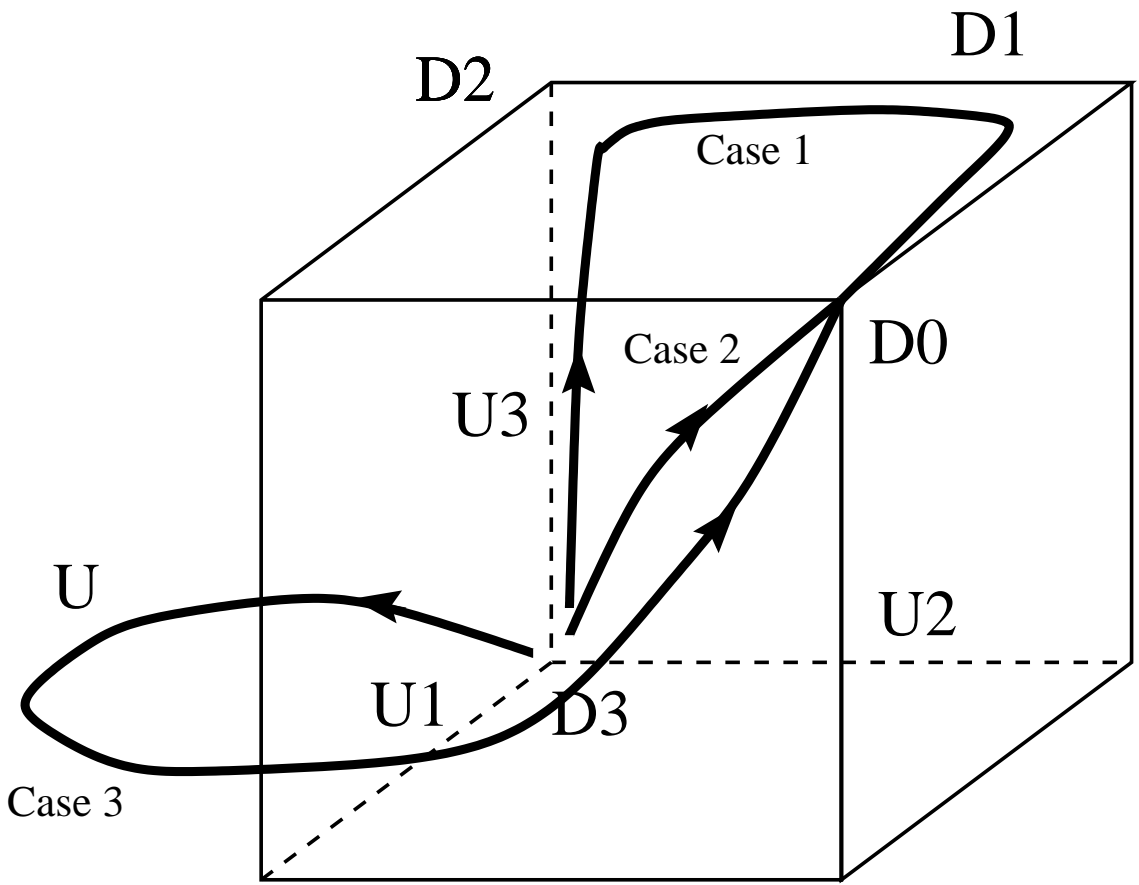}
\vspace{1.0cm}
\caption{RG phase diagram in the $u$-space for $m=3$,
$u=(u_1,u_2,u_3)$:
Case 1 represents a RG flow corresponding to the
successive transitions which are expected when
$u_3\gg u_2\gg u_1$.
Case 2 represents a direct transition to the completely reflecting
phase $D_0$, which happens when
$u_3\sim u_2\sim u_1$.
A direct transition to $D_0$ is also expected
for Case 3, where the Type (II) scattering potential plays
the role.}
\end{figure}

\end{document}